\documentclass[aip,apl,reprint]{revtex4-1}
\usepackage[utf8]{inputenc}
\usepackage{graphicx}
\usepackage[english]{babel} 
\usepackage{amsmath}
\usepackage{color}
\usepackage{psfrag}
\usepackage{float}
\usepackage{natbib}

\begin{document}
 
\title{Magneto-optical properties of InSb for terahertz applications}

\author{Jan Chochol}
\email{Jan.Chochol@vsb.cz, @dal.ca, @gmail.com}
\affiliation{Nanotechnology Centre, VSB -- Technical University of Ostrava, 17. listopadu 15/2172, 708 33 Ostrava – Poruba, Czech Republic}
\affiliation{Department of Electrical and Computer Engineering, Dalhousie University, 6299 South St, Halifax, NS B3H 4R2, Canada}

\author{Kamil Postava}
\affiliation{Department of Physics, VSB -- Technical University of Ostrava, 17. listopadu 15/2172, 708 33 Ostrava – Poruba, Czech Republic}

\author{Michael Čada}
\affiliation{Department of Electrical and Computer Engineering, Dalhousie University, 6299 South St, Halifax, NS B3H 4R2, Canada}

\author{Mathias Vanwolleghem}
\affiliation{Institut d'Electronique, de Microélectronique et de Nanotechnologie, UMR CNRS 8520, Avenue Poincaré, F-59652 Villeneuve d'Ascq cedex, France.}

\author{Lukáš Halagačka}
\affiliation{Nanotechnology Centre, VSB -- Technical University of Ostrava, 17. listopadu 15/2172, 708 33 Ostrava – Poruba, Czech Republic}
\affiliation{Department of Physics, VSB -- Technical University of Ostrava, 17. listopadu 15/2172, 708 33 Ostrava – Poruba, Czech Republic}

\author{Dominique Vignaud}
\affiliation{Institut d'Electronique, de Microélectronique et de Nanotechnologie, UMR CNRS 8520, Avenue Poincaré, F-59652 Villeneuve d'Ascq cedex, France.}

\author{Jean-François Lampin}
\affiliation{Institut d'Electronique, de Microélectronique et de Nanotechnologie, UMR CNRS 8520, Avenue Poincaré, F-59652 Villeneuve d'Ascq cedex, France.}

\author{Jaromír Pištora}
\affiliation{Nanotechnology Centre, VSB -- Technical University of Ostrava, 17. listopadu 15/2172, 708 33 Ostrava – Poruba, Czech Republic}

\date{\today}

\begin{abstract}
Magneto-optical permittivity tensor spectra of undoped InSb, n-doped and p-doped InSb crystals were determined using the terahertz time-domain spectroscopy (THz-TDS) and the Fourier transform far-infrared spectroscopy (far-FTIR). A Huge polar magneto-optical (MO) Kerr-effect (up to 20 degrees in rotation) and a simultaneous plasmonic behavior observed at low magnetic field (0.4 T) and room temperature are promising for terahertz nonreciprocal applications. We demonstrate the possibility of adjusting the the spectral rage with huge MO by increase in n-doping of InSb. Spectral response is modeled using generalized magneto-optical Drude-Lorentz theory, giving us precise values of free carrier mobility, density and effective mass consistent with electric Hall effect measurement.
\end{abstract}

\pacs{}

\maketitle 

Recent advances in the terahertz technology call for new devices and materials that would exhibit a non-reciprocal behavior in the terahertz range. One way to create the non-reciprocal behavior is the use of magneto-optical effects. Materials exhibiting a magneto-optical (MO) behavior at terahertz range are for example graphene \cite{tamagnone_near_2016, kuhne_invited_2014}, hexaferrites \cite{Shalaby:NatureCommun13} and semiconductors. General condition for successful applicability is the operation at low external magnetic field and at room temperature. Semiconductors are a viable choice, since their free carriers with low effective mass, either intrinsic or introduced by doping, allow for modulation of properties at these conditions.

The use of semiconductors in magneto-plasmonic devices for terahertz range has been suggested by Bolle et al. \cite{donald_m_bolle_application_1986}, while currently there are a number of studies, such as by Hu et al. \cite{Hu2012}, dealing with theoretical design of the devices. A correct implementation of the theoretical models is possible only when we know the exact properties of the materials used. Spectroscopic, non-destructive magneto-optical techniques give us the information we need.

The measurement of the free carrier magneto-optical effects in semiconductors has been called the ``Optical Hall effect'' by K\"uhne et al. \cite{kuhne_invited_2014} and Shubert et al. \cite{Schubert:03}, who developed a far infrared and terahertz ellipsometric, full Mueller matrix method for semiconductor characterization. The potential of the terahertz time-domain spectroscopy (THz-TDS) in investigation of semiconductors had been recognized by Mittleman \cite{mittleman_noncontact_1997}, (spatial inhomogeneities in GaAs), Jeon \cite{jeon_characterization_1998} (reflectivity of GaAs and Si), Grishowski \cite{grischkowsky_far-infrared_1990} (properties Si, Ge, GaAs) and Ino \cite{ino_terahertz_2004} (MO effect on InAs).

This letter deals with the magneto-plasmonic properties of InSb, since its low effective mass, 0.015 $m_0$ at $\Gamma$ point \cite{kittel_introduction_2004}, means its electrical and optical properties can be modulated by small magnetic fields. 

The area of spectroscopy of InSb in magnetic field has been pioneered by Lax et al. \cite{lax_cyclotron_1961}, while the subsequent theory had been summarized by Palik et al. \cite{palik_infrared_1970}, \cite{palik_coupled_1976} and in references therein, who studied semiconductors with reflective, transmittance and coupled mode measurements in the far infrared range, at low temperatures (liquid helium or nitrogen) and very high fields (several Tesla). The findings have also been reviewed by Pidgeon \cite{pidgeon1980free} and Kushwaha \cite{kushwaha_plasmons_2001}. Despite the wide range of semiconductors studied, the studies in the sixties and seventies were limited to the spectral range higher than $100 \rm{cm^{-1}}$ (3 THz) due to a lack of available sources. Consequently, the authors had to use higher magnetic fields and low temperatures to observe interesting magneto-plasmonic effects in their spectral ranges, which is inconvenient for practical applications. Our aim is to show that InSb can be used as a magneto-plasmonic material in terahertz range at room temperature and using low external magnetic fields.

We measured four InSb crystals from the manufacturer MTI Corp, polished wafers of 2'' diameter and $10\times10$ mm squares, 450 $\mu m$ thick, undoped and with doping n (Te) and p (Ge). The n-doped squares sample and the wafer have different concentrations, denoted $N_1$ and $N_2$ respectively.

We used two spectrometers to characterize the samples. The first one is the terahertz time-domain spectrometer TPS Spectra 3000 from TeraView Co., measuring in the THz range of 2-100 $\mathrm{cm^{-1}}$ (0.06-3 THz). The beam was focused using parabolic mirrors, through wire-grid polarizer to the sample at near normal incidence and reflects back through the same polarizer. The terahertz spectrometer measures both the amplitude and the phase of the reflected wave. Coupled with a model consistent with the Krames-Kroning relations, we can correctly obtain both real and imaginary parts of the permittivity.

The second one is the Fourier transform infrared spectrometer (FTIR) Bruker Vertex 70v, measuring in the far-infrared range of 50-680 $\mathrm{cm^{-1}}$ (1.5 - 20.4 THz), with the angle of incidence of 11 degrees and a variable analyzer and  polarizer azimuths.

\begin{figure}[htb]
	    \centering
	    \scriptsize{
  \psfrag{x}[c]{Wavenumber $\mathrm{cm^{-1}}$}
  \psfrag{0}[r]{0}
  \psfrag{0.2}[r]{0.2}
  \psfrag{0.4}[r]{0.4}
  \psfrag{0.6}[r]{0.6}
  \psfrag{0.8}[r]{0.8}
  \psfrag{1}[r]{1}
  \psfrag{100}[c]{100}
  \psfrag{200}[c]{200}
  \psfrag{300}[c]{300}
  \psfrag{400}[c]{400}
  \psfrag{500}[c]{500}
  \psfrag{600}[c]{600}
  \psfrag{700}[c]{700}
  \psfrag{ref}[c]{Reflectivity of all samples}
  \psfrag{dd}[l]{data}
  \psfrag{ff}[l]{fit}
  \psfrag{un}[l]{undoped}
  \psfrag{n1}[l]{$N_1$}
  \psfrag{n2}[l]{$N_2$}
  \psfrag{p}[l]{p} 
	    \includegraphics[width=.47\textwidth]{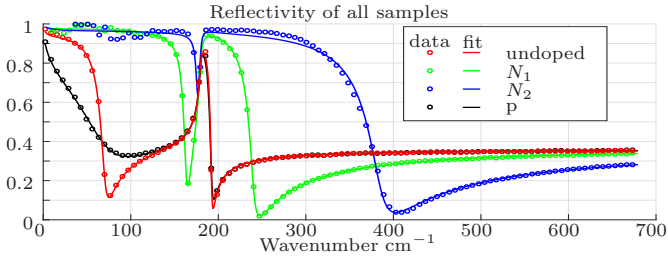}
	     }
	    \caption{Reflectivity of all InSb samples, TDS and FTIR data joined together.}
	    \label{fig:R_tog}
	  \end{figure}

All measurements were done in vacuum, in reflection, with a thick gold layer as a reference. The reflectivity of all measured samples is in Figure \ref{fig:R_tog}, the data in overlapping ranges were averaged.

The plasma edges, noticeable as the minima in the reflectivity are indicators of the level of doping/intrinsic concentration. The concentrations of carriers in undoped and p-doped InSb are roughly the same, but the p-doped sample has a much shorter scattering time, hence the shallow shape of the plasma reflectivity. The phononic peak arising from the lattice vibrations is visible at around 179 $\rm{cm^{-1}}$; different positions observed in the reflectivity are due to the effect of the free carriers.  

The the measured data are fitted to the following model. The permittivity in the THz and the far-infrared range is described using the Drude-Lorentz model, a sum of three parts,
\begin{equation}
\label{eq:D_L}     
  \varepsilon_{r}=\varepsilon_\infty-\underbrace{\frac{\omega_p^2}{\omega^2+i\gamma_p\omega}}_{\varepsilon_D}+\underbrace{\frac{A_L\omega_L^2}{\omega_L^2-\omega^2-i\gamma_L\omega}}_{\varepsilon_L}~,
\end{equation}
where the constant term $\varepsilon_\infty$ describes the background permittivity (high frequency absorbtions), the Drude term $\varepsilon_D$ describes the contribution of free carriers and the Lorentz term $\varepsilon_L$ comes from the lattice vibrations. In the Drude term, the plasma frequency is defined as
\begin{equation}
\omega_p=\left(\frac{Ne^2}{\varepsilon_0 m^*}\right)^\frac{1}{2},
\end{equation}
where $N$ is the carrier concentration, $e$ is the electron charge, $\varepsilon_0$ is the permittivity of free space, $m^*=m_{\rm{eff}} m_0$ is the effective mass of the charge carriers ($m_0$ is the mass of electron in vacuum) and $1/\gamma_p=\tau_p$ is the scattering time. The plasma frequency divided by $\sqrt{\varepsilon_\infty}$ is the frequency where the real part of permittivity crosses zero. The Lorentz term is characterized by the frequency $\omega_L$, the scattering time $\tau_L=1/\gamma_L$, and the amplitude $A_L$. 

\begin{figure}[htb]
	    \centering
	    \scriptsize{
	     \psfrag{x}[c]{Wavenumber $\mathrm{cm^{-1}}$}
  \psfrag{0}[r]{0}
  \psfrag{0.2}[r]{0.2}
  \psfrag{0.4}[r]{0.4}
  \psfrag{0.6}[r]{0.6}
  \psfrag{0.8}[r]{0.8}
  \psfrag{1}[r]{1}
  \psfrag{10}[c]{10}
  \psfrag{20}[c]{20}
  \psfrag{30}[c]{30}
  \psfrag{40}[c]{40}
  \psfrag{50}[c]{50}
  \psfrag{60}[c]{60}
  \psfrag{70}[c]{70}
  \psfrag{80}[c]{80}
  \psfrag{90}[c]{90}
  \psfrag{100}[c]{100}
  \psfrag{ref}[c]{Reflectivity (undoped)}
  \psfrag{ph}[l]{Phase (rad)}
  \psfrag{dd}[l]{data}
  \psfrag{ff}[l]{fit}
  \psfrag{T0}[l]{0 T}
  \psfrag{T4}[l]{0.43 T}
  \psfrag{T2}[l]{0.29 T}
	    \includegraphics[width=0.47\textwidth]{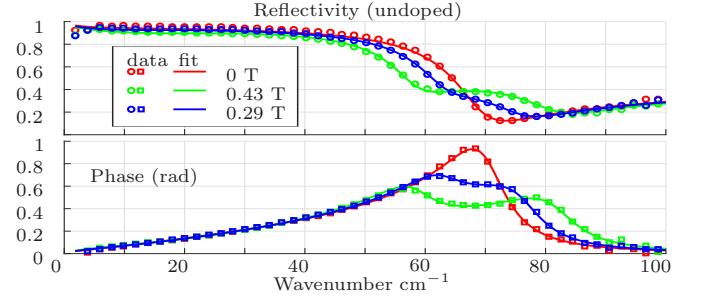}
	    }
	    \caption{TDS Reflectivity and corrected phase of undoped InSb in variable magnetic field}
	    \label{fig:R_un}
	  \end{figure}

When an external magnetic field is applied, the Drude term becomes anisotropic. The TDS reflectivity and phase of the undoped InSb in the variable magnetic field are shown in Figure \ref{fig:R_un}.

The reflectivity and phase of the samples is calculated using Berreman $4\times4$ method \cite{berreman_optics_1972}, allowing for a full anisotropy modeling. Reflection is modeled as a single interface between vacuum and a semiconductor with the permittivity $\varepsilon_r$.

The permittivity tensor used to describe the material with the magnetic field $B$ applied in the $z$ direction, perpendicular to the interface (MO polar configuration) is in the form
\begin{equation}
 \hat{\varepsilon_r}=\begin{bmatrix}
                \varepsilon_{xx} & \varepsilon_{xy} & 0 \\
                \varepsilon_{yx} & \varepsilon_{yy} & 0 \\
                0		 & 0                & \varepsilon_{zz}
               \end{bmatrix}.
\end{equation}
The $\varepsilon_{zz}$ component stays the same as $\varepsilon_r$ in (\ref{eq:D_L}) and $xx$, $yy$, $xy$, $yx$ components change to 
\begin{subequations}
 \label{eq:eps components reduced}
 \begin{align}
  \varepsilon_{xx}&=\varepsilon_{yy}=\varepsilon_\infty-\frac{\omega_p^2(\omega^2+i\gamma_p\omega)}{(\omega^2+i\gamma_p\omega)^2-\omega_c^2\omega^2}~+\varepsilon_L, \\
  \varepsilon_{xy}&=-\varepsilon_{yx}=-i\frac{\omega_p^2\omega_c\omega}{(\omega^2+i\gamma_p\omega)^2-\omega_c^2\omega^2}~,
 \end{align}
\end{subequations} 
 which contain an additional fitting parameter,  proportional to the magnetic field, the cyclotron frequency, defined as
 \begin{equation}
 \label{eq:cyclotron_fr}
  \omega_c=\frac{eB}{m^*}.
 \end{equation}
The Lorentz term can in theory be affected by the magnetic field, but the elements of the lattice are much heavier than electrons and the cyclotron frequency is negligible; the Lorentz term remains isotropic, as observed by K\"uhne at 8T in GaAs \cite{kuhne_invited_2014}.

The phase information in Figure \ref{fig:R_un} comes from three parts, $\varphi=\varphi_{\rm{sample}}-\varphi_{\rm{reference}}-\varphi_{\rm{shift}}$. $\varphi_{\rm{sample}}$ is the phase angle of the complex reflection coefficient of the sample and $\varphi_{\rm{shift}}$ stems from the misalignment $d$ of the sample and reference, as $\varphi_{\rm{shift}}= 4 d \pi \cos\alpha_i/\lambda$. The $\varphi_{\rm{shift}}$ is a fitting parameter in the data treatment ($d$ is on the order of 1-100 $\mu m$) and is subtracted from the data for plotting.

The magnetic field was created by a small permanent magnet with 0.43 T, with smaller fields obtainable through positioning of the magnet.

The measurements in different magnetic fields and no field were fitted together. The cyclotron frequency $\omega_c$ is 23.7 $\rm{cm^{-1}}$ for 0.43 T and the resulting effective mass of electrons in undoped InSb is $m_{eff}=(eB)/(\omega_c m_0)=0.0169$, which in accordance to theory \cite{zawadzki_electron_1974} is higher than frequently used value of 0.015. The knowledge of both the cyclotron frequency and the plasma frequency allows also for the calculation of the carrier concentration and mobility $\mu=e\tau_p/m^*$ and is necessary for the correct theoretical prediction of the behavior of magneto-plasmonic devices.

\begin{figure}[htb]
	    \centering
	    	    \scriptsize{
	     \psfrag{x}[c]{Wavenumber $\mathrm{cm^{-1}}$}
  \psfrag{0}[r]{0}
  \psfrag{0.1}[r]{0.1}
  \psfrag{0.2}[r]{0.2}
  \psfrag{0.3}[r]{0.3}
  \psfrag{0.4}[r]{0.4}
  \psfrag{0.5}[r]{0.5}  
  \psfrag{100}[c]{100}
  \psfrag{200}[c]{200}
  \psfrag{300}[c]{300}
  \psfrag{400}[c]{400}
  \psfrag{500}[c]{500}
  \psfrag{600}[c]{600}
  \psfrag{n1}[l]{$N_1$}
  \psfrag{n2}[l]{$N_2$}
  \psfrag{T0}[l]{0T}
  \psfrag{Tp}[l]{+0.43T}
  \psfrag{Tm}[l]{-0.43T}
  \psfrag{ref}[c]{Reflectivity}
\includegraphics[width=0.45\textwidth]{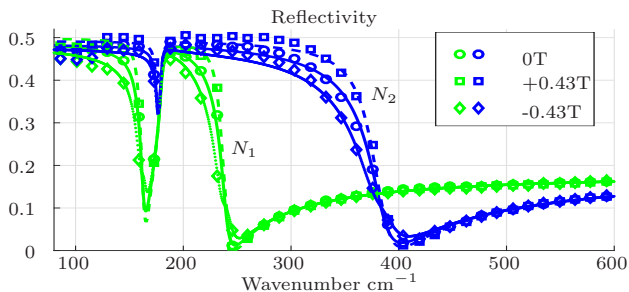}
    }
 \caption{Reflectivity of two concentrations of n-doped InSb, polarizer at 45. Data (symbols) and fit (curves) are compared. The center curve (circles, solid line) is reflectivity without the magnetic field, the other two are obtained for different signs of the magnetic field.}
\label{fig:R_n}
\end{figure}

The data from FTIR confirm the magneto-optical behavior of n-doped InSb, when the plasma frequency is pushed towards higher frequencies. Figure \ref{fig:R_n} shows the reflectivity of two n-doped samples when polarizer is at 45 degrees and analyzer at zero. When the direction of the magnetic field is reversed, the rotation of the reflected polarization changes direction, causing a drop/increase in reflected amplitude.

The parameters describing all samples are summarized in Table \ref{tab:Samples}. The non-magnetic properties of the undoped InSb match those reported or used by \cite{palik_indium_1997-1, Hu2012, gomez_rivas_optically_2006}, but our measurements also allow for calculation of correct effective mass and concentration. The samples have also been measured electrically by the van der Pauw (VDP) method \cite{l.j._van_der_pauw_method_1958}, which is equivalent to $\omega\rightarrow0$. The VDP data obtained are also listed in Table \ref{tab:Samples} and are in agreement to those obtained by spectroscopic measurement. The n-doped samples exhibit lower cyclotron frequency at the same magnetic field, meaning that the effective mass is higher, which is again in agreement with the theory \cite{zawadzki_electron_1974}.

\begin{table*}[htb]
\caption{Fitted and calculated parameters of the samples}
{%
\label{tab:Samples}
\begin{center}
\begin{tabular}{l|lllllll|lllllll}
Sample & $\omega_p/\sqrt{\epsilon_\infty}$ & $\tau_p $     & $\omega_L$                  & $\tau_L$       & $A_L$ & $\omega_c$ 			& $\varepsilon_\infty$ & $m_{eff}$ & $N_{spec.}$ 			   & $\mu_{spec.}$          & $\mathrm{N_{VDP}}$ 		    & $\mathrm{\mu_{VDP}}$              \\
       & $\left(\rm{cm^{-1}}\right) $      & $10^{-13}$ (s)& $\left(\rm{cm^{-1}}\right)$ & $10^{-12}$ (s) &       & $\left(\rm{cm^{-1}}\right)$ & 		       &           & $10^{17}$ $\left(\rm{cm^{-3}}\right)$ & $10^4 $ $(\rm{cm/Vs})$ & $10^{17}$ $\left(\rm{cm^{-3}}\right)$ & $10^4 \left(\rm{cm^{-3}}\right)$  \\\hline
und.   & 76.4 & 5.81 & 179.4 & 2.96 & 1.93 & 23.76 & 15.14 & 0.0169 & 0.17 & 5.77 & 0.20 & 6.66\\
$N_1$  & 217.9 & 4.75 & 179.4 & 2.96 & 2.13 & 11.18 & 15.58 & 0.0358 & 2.84 & 1.96 & 2.37 & 4.12\\
$nN_2$  & 378.1 & 1.98 & 179.9 & 2.95 & 2.07 & 13.93 & 15.12 & 0.0287 & 6.95 & 1.23 & - & -\\
p      & 84.1 & 0.75 & 179.4 & 2.96 & 1.94 & & 15.30 & - & - & - & 10.7 & 0.02 \\\hline
\end{tabular}
\end{center}
}%
\end{table*}

\begin{figure}[htb]
\centering
\scriptsize{
\psfrag{x}[c]{Wavenumber $\mathrm{cm^{-1}}$}
 \psfrag{-200}[l]{-200}
 \psfrag{-500}[l]{-500}
\psfrag{-1000}[l]{-1000}
 \psfrag{-1500}[l]{-1500}
 \psfrag{-2000}[l]{-2000}
  \psfrag{0}[l]{0}
  \psfrag{10}[l]{10}
    \psfrag{20}[l]{20}
    \psfrag{30}[l]{30}
      \psfrag{40}[l]{40}
  \psfrag{50}[l]{50}
 \psfrag{60}[l]{60}
  \psfrag{70}[l]{70}
\psfrag{80}[l]{80}
  \psfrag{100}[l]{100}
  \psfrag{150}[l]{150}
  \psfrag{200}[l]{200}
  \psfrag{300}[l]{300}
  \psfrag{400}[l]{400}
  \psfrag{500}[l]{500}
  \psfrag{600}[l]{600}
  \psfrag{nn0}[l]{$10^0$}
  \psfrag{nn2}[l]{$10^2$}
  \psfrag{nn5}[l]{$10^5$}
  \psfrag{rexx}[l]{$\Re\{\varepsilon_{xx}\}$}
  \psfrag{rexy}[l]{$\Re\{\varepsilon_{xy}\}$}
  \psfrag{imxx}[l]{$\Im\{\varepsilon_{xx}\}$}
  \psfrag{imxy}[l]{$\Im\{\varepsilon_{xy}\}$}
  \psfrag{un}[l]{und.}
  \psfrag{n1}[l]{$N_1$}
  \psfrag{n2}[l]{$N_2$}
  \psfrag{p}[l]{p}
  \psfrag{0T}[l]{(0T)}
  \psfrag{2T}[l]{(0.29T)}
  \psfrag{4T}[l]{(0.43T)}
  
\includegraphics[width=0.47\textwidth]{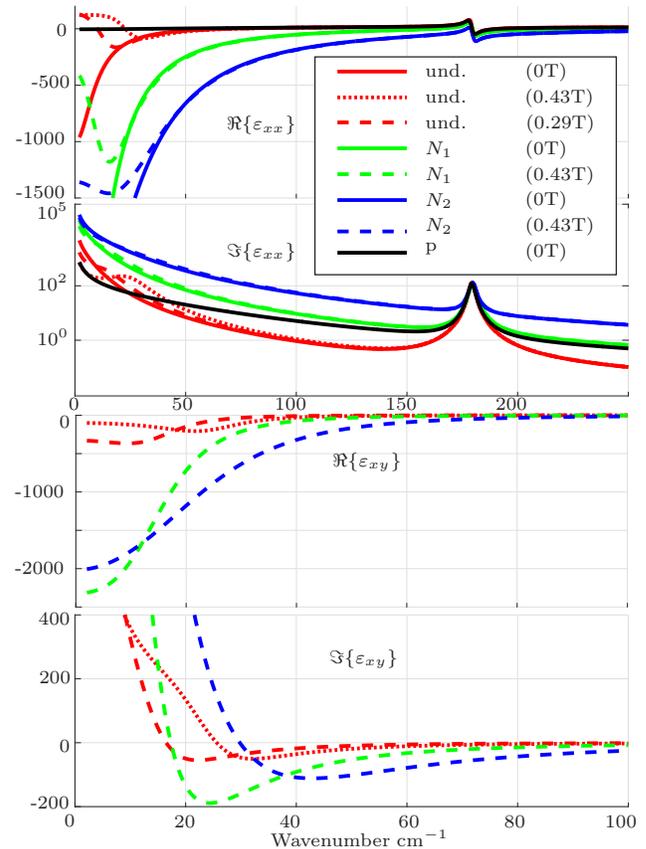}
}
\caption{Calculated diagonal and off-diagonal complex permittivity of all samples of InSb with and without applied magnetic field. Note the different ranges/scales to highlight important features.}
\label{fig:per_tog}
\end{figure}

Figure \ref{fig:per_tog} show the obtained permittivity of all the samples using fitted parameters listed in Table \ref{tab:Samples}.  The presence of a cyclotron frequency changes the low frequency limit of the real part of the diagonal components $\varepsilon_{xx,yy}$, which can completely change sign, if the cyclotron frequency is high enough. The magnetic field also increases absorbtions (Landau level absorbtion) at $\omega_c$, noticeable mainly in the undoped InSb sample. This effect is usually observed when $\omega_c>\omega_p$ and also causes changes in the effective mass, which is negligible in our case due to low magnetic field \cite{palik_free_1961}. 
The off-diagonal elements exhibit a peak around $\omega_c$ and a limit at low frequency (the classical Hall effect), while the imaginary part goes to infinity for low-frequencies \cite{yu_fundamentals_2013}.

\begin{figure}[htb]
  \centering
\scriptsize{
  \psfrag{x}[c]{Wavenumber $\mathrm{cm^{-1}}$}
  \psfrag{y}[c]{Degrees}
  \psfrag{0}[r]{0}
  \psfrag{10}[r]{10}
  \psfrag{20}[r]{20}
  \psfrag{30}[r]{30}
  \psfrag{-10}[r]{10}
  \psfrag{-20}[r]{20}
  \psfrag{-30}[r]{30}
  \psfrag{100}[c]{100}
  \psfrag{200}[c]{200}
  \psfrag{300}[c]{300}
  \psfrag{400}[c]{400}
  \psfrag{500}[c]{500}
  \psfrag{unt}[l]{und. $\theta$}
  \psfrag{une}[l]{und. $\epsilon$}
  \psfrag{n1t}[l]{$N_1$ $\theta$}
  \psfrag{n1e}[l]{$N_1$ $\epsilon$}
  \psfrag{n2t}[l]{$N_2$ $\theta$}
  \psfrag{n2e}[l]{$N_2$ $\epsilon$}
  \includegraphics[width=0.47\textwidth]{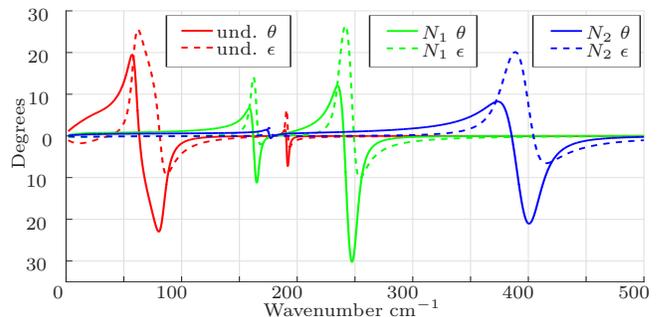}
}
 \caption{Obtained polar Kerr rotation $\theta$ and ellipticity $\epsilon$}
\label{fig:Kerr_tog}
\end{figure}

Using the permittivity tensors we can further obtain the Kerr effect which is a good metric to describe the magneto-optical behavior of materials. The polar magneto-optical Kerr effect is a change of the polarization ellipse azimuth and ellipticity upon reflection of linearly polarized light from a sample in a magnetic field perpendicular to the interface \cite{visnovsky_optics_2006}. Figure \ref{fig:Kerr_tog} shows the obtained rotation and ellipticity.

The Figures \ref{fig:per_tog} and \ref{fig:Kerr_tog} give us the idea of the applicability of InSb as a magneto-plasmonic material. The magneto-optical effects is strongest around sharp changes in the original permittivity, either around the plasma edge or the lattice vibration. There are regions, where the materials exhibit a strong Kerr rotation, a small Kerr ellipticity while the $\epsilon_{zz}$ component remains plasmonic, for undoped InSb its bellow 50 $\rm{cm^{-1}}$ (1.5 THz). For the n-doped samples, the behavior is similar, only shifted towards higher frequencies. This means that even though increasing carrier concentration increases effective mass and therefore lowers the cyclotron frequency, a strong magneto-plasmonic behavior is still present, allowing for a fine-tuning of the material and device properties. The p-doped sample didn't exhibit any measurable magneto-optical activity, due to very low cyclotron frequency caused by the effective mass of the heavy holes.

In conclusion, we have presented magneto-optical measurement of four samples of InSb with different carriers and carrier concentrations. The data are in good agreement with the Drude-Lorentz model in external magnetic field, both in the amplitude and phase, ensuring a valid use of the Kramers-Kronig relations. The materials exhibit a large magneto-plasmonic activity for the n-type carriers around the plasma edge, for all three levels of free electron concentration. The polar Kerr rotation is up to 20 degrees, which is huge compared to the typical values of tenths of degrees in visible range. The strength of this effect at room temperature and reasonably low magnetic field points to applicability of InSb as a material for non-reciprocal magneto-plasmonic devices usable in the terahertz range, plus the properties can easily be further modulated by either heat \cite{gomez_rivas_low-frequency_2006} or light \cite{gomez_rivas_optically_2006}, or the material can be used in a heterostructure in combination with different doping levels. The plasmonic properties can further enhance the magneto-optical effects by capturing, guiding and concentrating light at subwavelength scale using surface plasmons.  Moreover, the physical properties obtained by spectroscopic measurement agree with electrical measurement and give us a correct value of the effective mass for further use in non-reciprocal device design.

\begin{acknowledgments}
 This work was supported in part by projects GA15-08971S, ``IT4Innovations excellence in science - LQ1602'', SGS project SV 7306631/2101, CREATE ASPIRE Program supported by NSERC and research grant JCJC TENOR ANR-14-CE26-0006. 
\end{acknowledgments} 

\bibliography{MyLibrary_manual}
\end{document}